\documentclass[12pt, a4paper]{article}

\usepackage[
  margin=0.75in,
  headsep=10pt, 
]{geometry}

\usepackage{graphicx}
\usepackage{soul}
\usepackage{graphicx}
\usepackage{epstopdf, epsfig}
\usepackage{amsmath}
\usepackage[table]{xcolor}
\usepackage{subcaption}
\usepackage{soul}
\usepackage{makecell}
\usepackage{multirow}
\usepackage{breqn}
\usepackage{amssymb} 
\usepackage{booktabs}
\usepackage{dcolumn}
\usepackage{bm}
\usepackage[utf8]{inputenc}
\usepackage[T1]{fontenc}
\usepackage{mathptmx}
\usepackage{etoolbox}
\usepackage{hyperref}
\usepackage{subcaption}
\usepackage{ragged2e}
\usepackage[sort&compress]{natbib} 

\newcommand{\edt}[1]{{\color{black}#1}} 
\newcommand{\mk}[1]{{\color{black}#1}} 

\newcommand{\soptitle}{On the role of morphology and kinematics of biological swimmers to spread and suppress their odors in the wake}

\usepackage{xcolor} 

\begin{document}


\begin{center}
\Large \bf{\soptitle}
\vspace{0.1in}
\end{center}

\begin{center}
{Maham Kamran$^1$, Amirhossein Fardi$^1$, Chengyu Li$^2$, and Muhammad Saif Ullah Khalid$^{1\star}$}\\
\vspace{0.1in}
\end{center}
\begin{center}
$^1$Nature-Inspired Engineering Research Lab (NIERL), Department of Mechanical \edt{and} Mechatronics Engineering, Lakehead University, Thunder Bay, ON P7B5E1, Canada\\
$^2$Department of Mechanical and Aerospace Engineering, Case Western Reserve University, Cleveland, OH 44106, USA\\
\vspace{0.05in}
$^\star$\small{Corresponding Author, Email: mkhalid7@lakeheadu.ca}
\end{center} 

\begin{abstract}
\mk{Understanding the interplay between hydrodynamics and chemical sensing in aquatic environments is crucial for unraveling biological swimmers' navigation, foraging, and communication strategies. This study investigates the role of kinematics and morphologies of fish in dispersion and suppression of odor cues in their wake. We employ high-fidelity three-dimensional computational fluid dynamics simulations, integrating a sharp-interface immersed-boundary method with an odor transport model. Using carangiform and anguilliform kinematics for a jackfish and an eel, we analyze the transport of chemical cues in the wake of undulatory swimmers at a Reynolds number of $3000$ and Strouhal numbers of $0.25$ and $0.4$. Our findings reveal that odor plumes closely align with vortex structures, emphasizing a strong coupling between hydrodynamics and chemical dispersion. We demonstrate that kinematics, rather than morphology, predominantly govern odor transport, with anguilliform motion generating broader, more persistent odor trails. Increasing the amplitude of undulation improves the effectiveness of the odor, driven primarily by convection, while diffusion plays a secondary role. These insights provide a deeper understanding of underwater sensing mechanisms and inform the design of bio-inspired robotic systems with improved navigation and chemical detection capabilities.}
\end{abstract}

\section{Introduction}
\label{sec:Intro}
Nature’s remarkable designs inspire researchers seeking effective strategies to address complex problems. Among these, the ability of fish to perform highly efficient propulsion and navigation in aquatic environments captivated scientists and engineers alike. With advancements in robotic technologies for applications\edt{,} such as environmental monitoring, \edt{underwater} search, rescue, and exploration, understanding fish biomechanics and hydrodynamics \edt{becomes} pivotal in designing bio-inspired underwater robots. Fish, as exceptional swimmers, achieve efficient propulsion and maneuverability through intricate body flexion and undulatory kinematics, which inspired numerous studies to investigate their hydrodynamic performance \cite{triantafyllou2000hydrodynamics, sfakiotakis1999review, lauder2005hydrodynamics, fish2017control}.

Early research focused on understanding \edt{propulsive} mechanisms and maneuverability by manipulating parameters\edt{,} such as undulation frequency and wavelength \cite{khalid2020flow, zhang2022vortex, triantafyllou2000hydrodynamics}. For instance, studies revealed that larger wavelengths \edt{enhanced} thrust for carangiform swimmers, enabling them to conserve energy at optimal wavelengths while maintaining high speeds \cite{anderson2001boundary, khalid2021larger, videler1993fish}. Conversely, anguilliform swimmers \edt{were found to achieve} superior hydrodynamic performance at smaller wavelengths relative to their body length \cite{khalid2021anguilliform, gazzola2014scaling}. These findings highlight the diverse strategies employed by marine animals, fostering the development of bio-inspired underwater vehicles with enhanced propulsion capabilities \cite{fish2006passive, fish2020bio, lauder2015fish, zhang2022physical}. While significant progress \edt{was} made in understanding fish locomotion \edt{in the previous two decades}, the role of chemical cues and their interaction with hydrodynamics remains \edt{largely} an \edt{unexplored} area of \edt{research} \cite{cox2008hydrodynamic, bronmark2012aquatic, montgomery2014sensory, carde2021navigation}. In underwater environments, fish rely on an array of sensory modalities—such as vision, acoustic perception, lateral line sensing, and olfaction for navigation, foraging, and communication \cite{montgomery2014sensory, bleckmann2014central, bleckmann2001lateral}. Olfaction, \edt{involved in} detecting chemical cues like amino acids, pheromones, and metabolic by-products, \edt{particularly} plays a crucial role in guiding behaviors related to migration, foraging, and predator avoidance \cite{bunnell2011fecal, domenici2011animal, carde2021navigation}. Despite \edt{being an important biological underwater sensing technique}, the coupling between odor dynamics and vortex dynamics in aquatic environments received \edt{very} limited attention in the literature \cite{kamran2024does, lei2023numerical, lei2023wings}. 

\edt{With this context,} our \edt{recent} work explored the coupling between vortex dynamics and odor transport using a two-dimensional (2D) \edt{simulation} framework \cite{kamran2024does}. Through \edt{our} in-house \edt{i}mmersed-\edt{b}oundary \edt{m}ethod (IBM) \edt{based} solver, we analyzed how carangiform and anguilliform swimmers \edt{influenced} odor dispersion in water and air. The study revealed that vortex dynamics significantly \edt{affected the trajectory of odor spots}, with convection dominating in water and diffusion playing a more prominent role in air. Additionally, we observed a phase difference between the vortex structures and the odor \edt{cues}, indicating that the alignment between coherent flow patterns and odor spots \edt{was} not always precise. Anguilliform swimmers were found to generate stronger and more dispersed chemical cues compared to carangiform swimmers. \edt{Our present work} extends \edt{this} analysis to \edt{three-dimensional (3D)} flows \edt{and odor transport in the wake of marine swimmers}, enabling a more realistic representation of \edt{swimmers'} morphology and kinematics. Unlike 2D \edt{computational modeling}, 3D simulations capture complex flow phenomena\edt{,} such as cross-flow interactions and 3D vortex structures, providing deeper insights into the transport of chemical cues and their coupling with hydrodynamics. Furthermore, the incorporation of realistic geometries and kinematics of biological swimmers, such as \edt{j}ackfish and \edt{e}els, allows us to investigate how physiological features and motion patterns influence odor dispersal and suppression in aquatic environments. The primary aim of this work is to address the following research questions: (i) \edt{h}ow do integrated vortex-odor dynamics in the wake of fish influence underwater sensing? and (ii) \edt{h}ow do physiological \edt{features} and kinematic characteristics contribute to the transport of chemical cues?

To answer these questions, we employ \edt{our} in-house computational framework based on the sharp-interface \edt{IBM} \cite{farooq4874977accurate}. This solver, named \edt{VortexDyn}, is designed to simulate fluid-structure-chemical interactions with high accuracy. By integrating odor transport equations \edt{along with} the 3D Navier-Stokes \edt{formulations}, we capture the unsteady coupling between hydrodynamic \edt{flow field} and \edt{dispersal of chemical cues}. Our simulations focus on transitional flow regimes with a Reynolds number \edt{(\mbox{Re}) of $3,000$} and Strouhal numbers ($\mbox{St} = 0.25$ and $0.4$), combined with Schmidt number ($\mbox{Sc}$) to characterize odor transport in water. These investigations offer valuable insights into how \edt{swimmers'} kinematics, flow conditions, and \edt{properties of the fluidic medium} influence the distribution of chemical \edt{odor in the wake}. Understanding the relationship between vortex dynamics and odor transport has significant implications for underwater sensing and robotics. By mimicking the natural strategies of fish, bio-inspired robotic systems can integrate hydrodynamic and chemical sensing capabilities to perform complex tasks in challenging aquatic environments. This work \edt{significantly} advances \edt{our knowledge} by addressing the role of $\mbox{3D}$ vortex dynamics in odor dispersal and offering a comprehensive simulation framework \edt{for fluid-structure-odor interactions}. Our findings are expected to contribute to the fields of marine biology, environmental conservation, and oceanographic exploration, while also aiding in the development of efficient and adaptive underwater robotic systems \cite{bianchi2021bio, marras2012fish}.

\section{Computational Methodology}
\label{sec:Num_Method}
\begin{figure}[htbp]
{\includegraphics[width=1\textwidth]{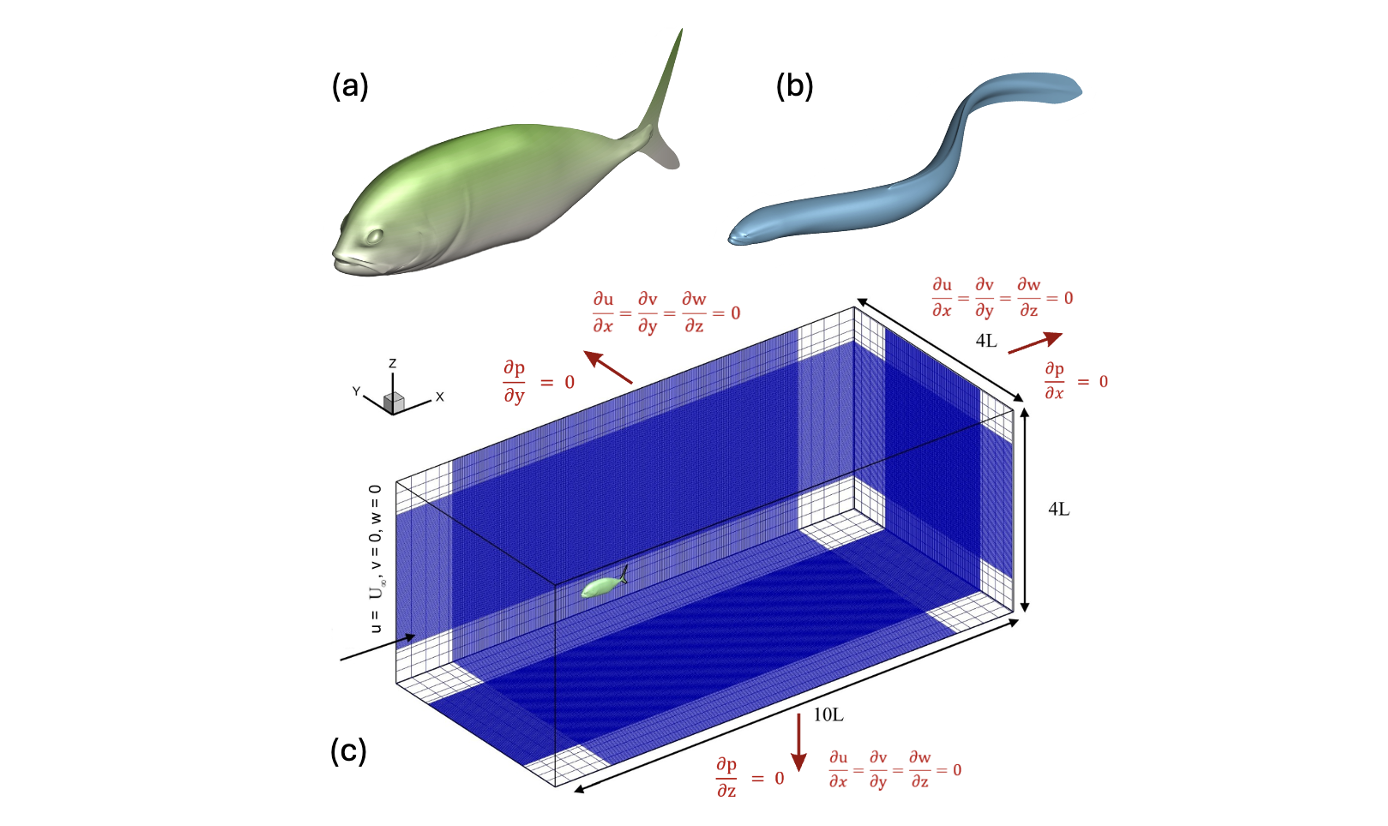}}
\caption{ Physiological model of (a) Jackfish, (b) Eel, and (c) Virtual tunnel, non-uniform Cartesian grid, and the boundary conditions for simulating flows }
\label{fig:flow_domain}
\end{figure}

In this study, we use the \edt{real} geometries of \edt{a jackfish} \cite{khalid2021larger} and \edt{an eel} \cite{khalid2021anguilliform}, as shown in Figs.~\ref{fig:flow_domain}a and \ref{fig:flow_domain}b, to represent two wavy kinematic modes: carangiform and anguilliform, respectively. Carangiform swimmers, such as \edt{j}ackfish, possess prominent caudal fins attached to their bodies \cite{lauder2005hydrodynamics}, while anguilliform swimmers, like \edt{e}els, undulate a large portion of their \edt{slender} bodies to generate movement. \edt{In this work}, the total height and width of the \edt{j}ackfish are $0.3021L$ and $0.1481L$, respectively, where $L$ is the total length of the fish. The area of \edt{its body} is $0.4479L^2$. It is important to note that median fins, such as the dorsal and anal fins, are not considered in this study, as their total effect is less than $5\%$ of the forces \cite{Liu_Ren_Dong_Akanyeti_Liao_Lauder_2017}. The total height and width of the \edt{eel} are $0.0884L$ and $0.0605L$, respectively, \edt{whereas its body has a surface area of} $0.1739L^2$. The \edt{geometry of the jackfish} is comprised of the trunk and caudal fin \cite{khalid2021larger}, where \edt{the trunk is} discretized using 22,712 triangular elements and 11,358 nodes, and the caudal fin alone consists of 2,560 triangular elements and 1,369 nodes. The \edt{eel's} geometry is represented using 33,112 triangular elements and 16,558 nodes \cite{khalid2021anguilliform}.

Each swimmer is positioned in a virtual tunnel to simulate flows over it, with the flow domain dimensions illustrated in Fig.~\ref{fig:flow_domain}c. The dimensions of the flow domain are set to ${10L} \times {4L} \times {4L}$, where $L$ represents the total length of the swimmer.  We use the same approach as Kamran et al. \cite{kamran2024does} to design the non-uniform mesh with two distinct regions, as shown in Fig.~\ref{fig:flow_domain}c: a finer region and a coarser region.


The amplitude of the carangiform \edt{kinematic} profile is expressed by the following equation \cite{khalid2016hydrodynamics, khalid2020flow, khalid2021larger, kamran2024does}:

\begin{align}
    A\left(\frac{x}{L}\right) &= 0.02 - 0.0825\left(\frac{x}{L}\right) + 0.1625\left(\frac{x}{L}\right)^2; \quad 0 < \frac{x}{L} < 1
\label{eq:carangiform}
\end{align}

\noindent \edt{where}, $x$ represents the stream-wise coordinate of each node used to discretize the 3D model swimmer, and $A(x/L)$ denotes the local amplitude at a specific position along the swimmer's body, nondimensionalized by its total length ($L$). The coefficients \edt{here} are derived from data for a steadily swimming \edt{saithe} fish, a carangiform swimmer \cite{videler1993fish}, with local amplitudes of $A(0)=0.02$, $A(0.2)=0.01$, and $A(1.0)=0.10$. \edt{Previously, Khalid et al. \cite{khalid2021larger} clearly showed that this kinematic profile closely matched with a real jackfish.} To ensure that the maximum amplitude at the trailing edge does not exceed $0.10$ for both swimmers, we define the amplitude envelope of the anguilliform kinematics using the following relation \cite{maertens2017optimal, khalid2020flow, khalid2021anguilliform}:

\begin{align}
    A\left(\frac{x}{L}\right) &= 0.0367 + 0.0323\left(\frac{x}{L}\right) + 0.0310\left(\frac{x}{L}\right)^2; \quad 0 < \frac{x}{L} < 1
\label{eq:amp}
\end{align}

Using $f$ as the oscillation frequency, \edt{$\lambda$ as the wavelength,} and $t$ as the time, we model the undulatory motion for both cases with the following equation:

\begin{align}
    y\left(\frac{x}{L}\right) &= A\left(\frac{x}{L}\right) \sin\left[2\pi\left(\frac{x}{\lambda} - ft\right)\right]
\label{eq:motion}
\end{align}
 

The \edt{governing} mathematical model for fluid \edt{flows} is based on the following non-dimensional forms of the continuity and incompressible Navier-Stokes equations \cite{mittal2008versatile, farooq4874977accurate}:

\begin{align}
    \frac{\partial u_j}{\partial x_j} &= 0,\\
    \frac{\partial u_i}{\partial t} + {u_j} \frac{\partial u_i}{\partial x_j} &= -\frac{1}{\rho} \frac{\partial p}{\partial x_i} + \frac{1}{Re} \frac{\partial^2 u_i}{\partial x_j \partial x_j} + f_b,
\label{eqn:CartCont}
\end{align}

\noindent where $\{i, j\} = \{1, 2, 3\}$, $x_i$ represents the Cartesian directions, $u_i$ denotes the Cartesian components of the fluid velocity, $p$ is the pressure, and $\mbox{Re}$ is the Reynolds number. \edt{We define it as $\mbox{Re}={U_\inf}{L}/\nu$, where $U_\infty$ shows the free-stream velocity, and $\nu$ is kinematic viscosity of the fluid.} In this formulation, $f_b$ is a discrete forcing term that enables a sharp representation of the immersed boundary \cite{mittal2008versatile, farooq4874977accurate}. We solve \edt{these} governing equations using a sharp-interface \edt{IBM} on a non-uniform Cartesian grid. Radial basis functions serve as the interpolation scheme to precisely identify the immersed bodies \cite{farooq4874977accurate}. A central difference scheme is employed for spatial discretization of the diffusion term, while the Quadratic Upstream Interpolation for Convective Kinematics (QUICK) scheme is used for the convection term. Temporal integration is performed using a fractional-step method, achieving second-order accuracy in both time and space. The prescribed wavy kinematics are applied as a boundary condition on the swimmer's body through a ghost-cell technique \cite{farooq4874977accurate}, which accommodates both rigid and flexible structures. Details of this fully parallelized solver and its application to various bio-inspired \edt{fluid-structure interactions} related problems are provided by Farooq et al. \cite{farooq4874977accurate}. Neumann boundary conditions are applied at the far-field boundaries, except at the left-side inlet boundary, where Dirichlet conditions specify the inflow. 



Chemical sensing \edt{plays} a crucial role in navigation, foraging, and communication by diffusion \edt{and advection} of chemical cues released in the fluid\edt{, surrounding a swimmer}. These cues consist of molecules such as amino acids, pheromones, and metabolic waste products. \edt{In this work, we model dynamics of odor using the chemical transport equation} to determine the instantaneous odor concentration \edt{fields} in the entire computational domain. The governing equation for the unsteady convection-diffusion of the odorant is as follows:

\begin{align}
    \frac{\partial C}{\partial t} + {u_i} \frac{\partial C}{\partial x_i} &= D \frac{\partial^2 C}{\partial x_i \partial x_i},
\label{eqn:odor_transport}
\end{align}

\noindent where, $C$ represents the odor concentration, nondimensionalized by the source concentration at the swimmer's body surface, and $D$, denotes the diffusivity of the odor. The temporal term (first term on the left-hand side), the convective term (second term on the left-hand side), and the diffusion term (on the right-hand side) in Eq.~\ref{eqn:odor_transport} are discretized using the same computational schemes applied to the Navier-Stokes equations (Eq.~\ref{eqn:CartCont}). This robust computational framework enables multi-physics simulations involving fluid-structure-chemical interactions, allowing the accurate modeling of odor transport even at very high Schmidt numbers. The interplay of convection and diffusion processes is analyzed to describe transport of odor within the flow field, explaining \edt{important physical} mechanisms \edt{that drive underwater} chemical sensing for \edt{biological marine species}.

Before conducting the multi-physics simulations in this study, we perform extensive grid-independence and time-step convergence studies. For this purpose, we use the geometry of \edt{a jackfish} with prescribed carangiform kinematics \cite{khalid2020flow}, characterized by an undulation wavelength ($\lambda/L$) of $1.05$. The measured Strouhal number ($\mbox{St}$ or $f$*) is $\mbox{St}=0.40$, where $\mbox{St} = {2}\edt{{A_\circ}}{f}/U_\infty$. Here, $f$ denotes the excitation or flapping frequency of the caudal fin, \edt{and} \edt{$A_\circ$} is the maximum one-sided oscillation amplitude of the caudal fin (a measure of \edt{the} wake width).

\begin{figure}[htbp]
    \centering
    \begin{minipage}{0.5\textwidth}
        \centering
        \includegraphics[width=\linewidth]{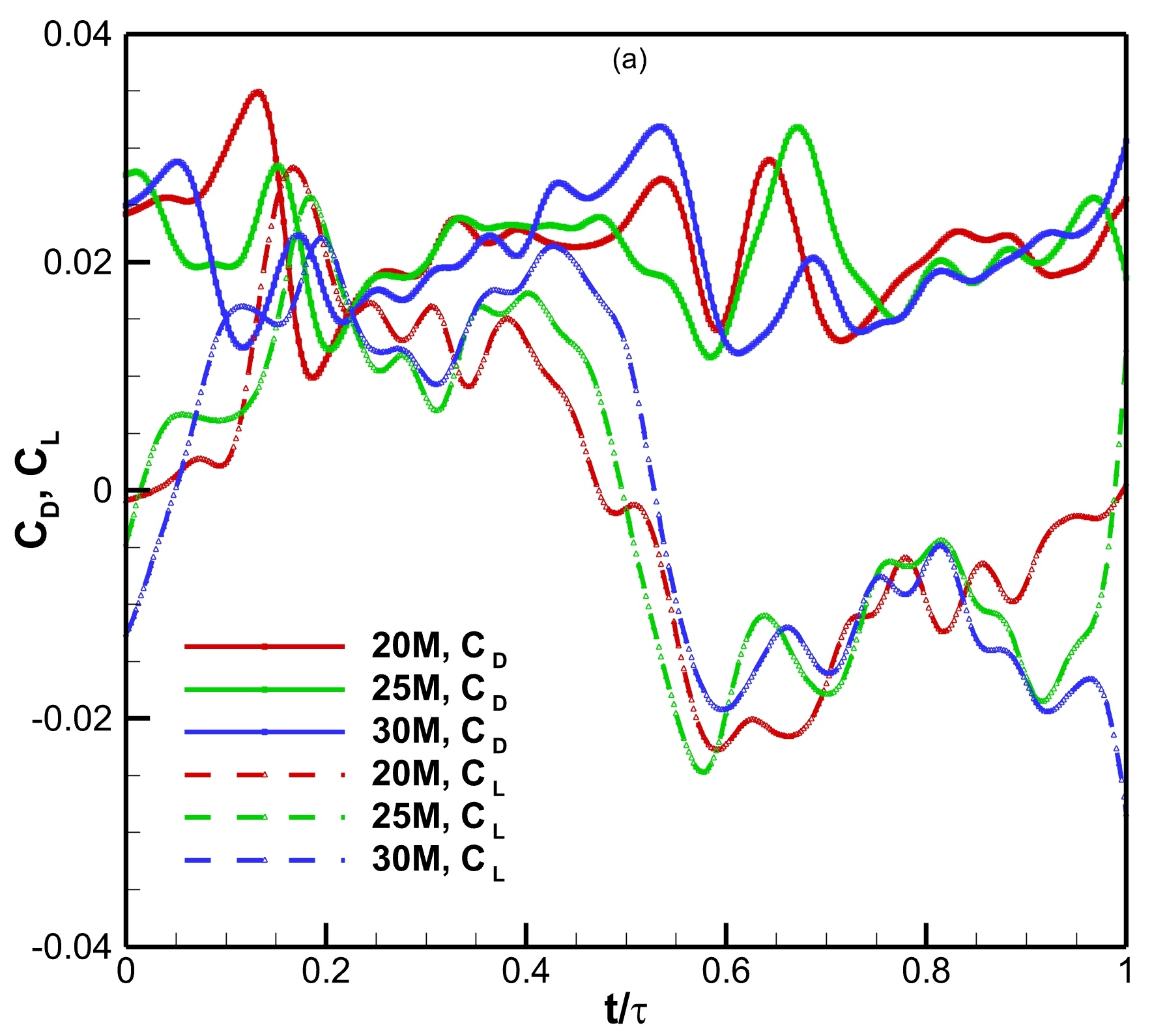}
    \end{minipage}%
    \begin{minipage}{0.5\textwidth}
        \centering
        \includegraphics[width=\linewidth]{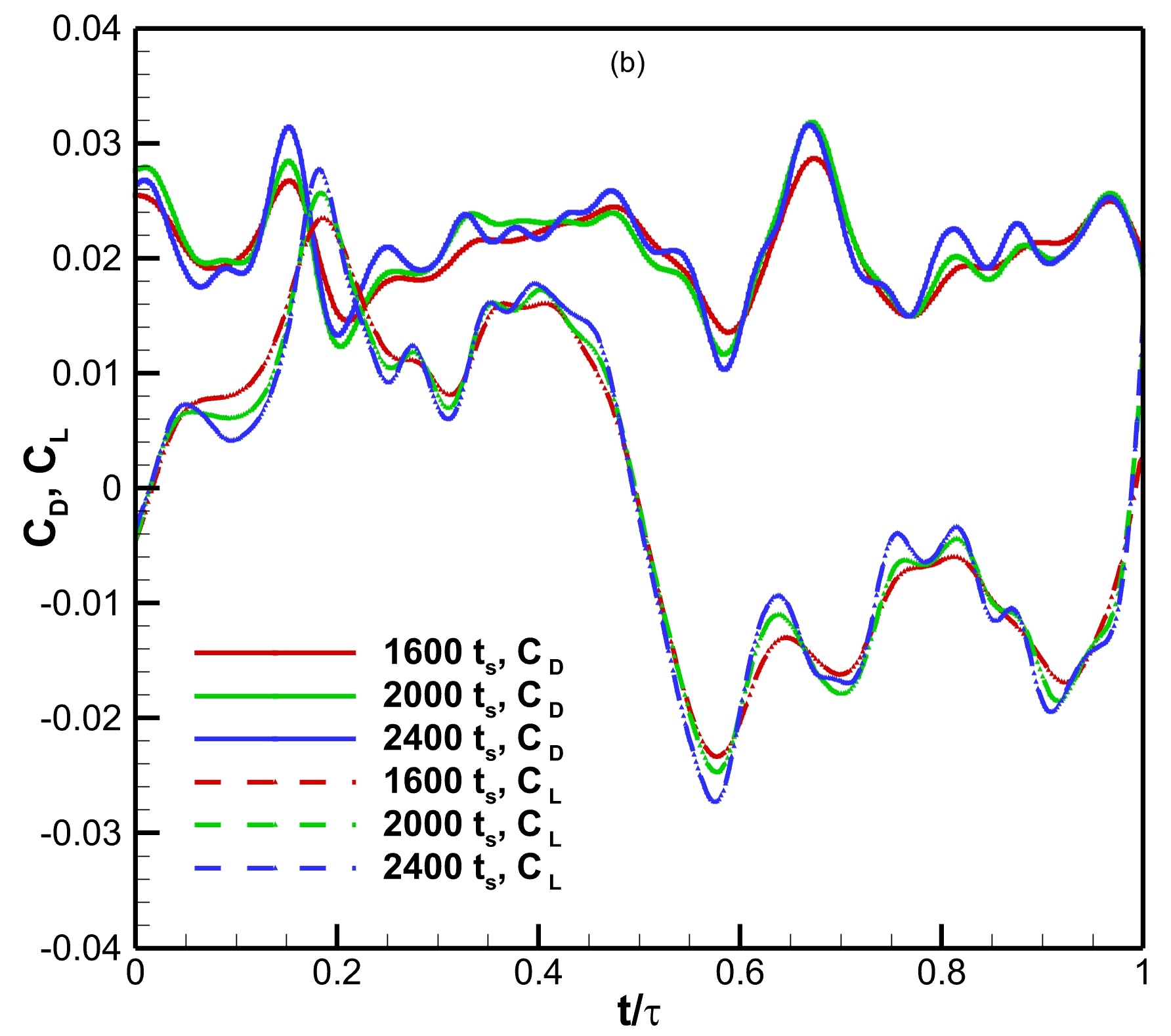}
    \end{minipage}
    \caption{\small Results for convergence of (a) grid size and (b) time step size}
    \label{Results for convergence}
\end{figure}

The grid-independence test is conducted using three grid configurations and a time step (${\Delta}t$) corresponding to $2000$ time steps per oscillation cycle. The grid sizes are defined as $505\times209\times209$ \edt{(Grid $1$)}, $553\times217\times217$ \edt{(Grid $2$)}, and $601\times225\times225$ \edt{(Grid $3$)}, corresponding to approximately $20$ million, $25$ million, and $30$ million mesh \edt{nodes}, respectively. Figure~\ref{Results for convergence}a shows the comparison of instantaneous lift and drag coefficients ($C_L$ and $C_D$) for one complete undulation cycle, where $\tau$ is the time period of the undulation. The force coefficient profiles for the three grids are qualitatively similar during the $15^{\text{th}}$ cycle after steady-state conditions are achieved within $5$–$6$ oscillation cycles.

To evaluate grid convergence and select an appropriate grid for further simulations, we calculate the standard error of estimate ($SE_e$), defined as \cite{kamran2024does}:

\begin{equation}
SE_e = \sqrt{\frac{\sum (Y_{\text{obs}} - Y_{\text{pred}})^2}{n - p}},
\label{eqn:standarderror}
\end{equation}

\noindent where \( Y_{\text{obs}} \) represents the observed values, \( Y_{\text{pred}} \) represents the predicted values, \( n \) is the number of observations, and \( p \) is the number of parameters. Using the results from Grid $3$ as reference ($Y_{\text{pred}}$), Table~\ref{Comparison of Grid and Time-step independence study results} shows smaller differences between the results obtained from Grid $2$ and Grid $3$ for both $C_L$ and $C_D$. Consequently, we select Grid $2$ with $25$ million \edt{nodes} for subsequent simulations.

For the time step convergence study, we test three time step sizes, $\Delta{t}_1$, $\Delta{t}_2$, and $\Delta{t}_3$, corresponding to $1600$, $2000$, and $2400$ time steps per oscillation cycle. The results of $C_L$ and $C_D$ are shown in Fig.~\ref{Results for convergence}b \edt{for the three simulations}. The force coefficient profiles are \edt{very} similar \edt{for} all three time steps. Using the smallest time step size ($\Delta{t}_3$) as the reference, Table~\ref{Comparison of Grid and Time-step independence study results} illustrates that the differences in results are smaller between $\Delta{t}_2$ and $\Delta{t}_3$ than \edt{that} between $\Delta{t}_1$ and $\Delta{t}_3$. Therefore, the remaining simulations in this work are performed using $2000$ time steps per undulation cycle. For validation of our computational methodology, we refer the reader to the recent \edt{studies} of Farooq et al. \cite{farooq4874977accurate} and Kamran et al. \cite{kamran2024does}.

\begin{table}[htbp]
\centering 
\caption{\small Comparison of $SE_e$ for grid-convergence and time step independence tests}
\vspace{-25pt}
\resizebox{\textwidth}{!}{%
\small 
\begin{tabular}{@{}cccccccc@{}}
\multicolumn{1}{l}{} & \multicolumn{1}{l}{} & \multicolumn{1}{l}{} & \multicolumn{1}{l}{} & \multicolumn{1}{l}{} \\ \toprule\toprule
& \multicolumn{2}{c}{Grid-independence Tests}  &&  \multicolumn{2}{c}{Time step Convergence Tests} \\ \cline{2-3} \cline{5-6}
\textbf{} & Grid 1 \& Grid 3 & Grid 2 \& Grid 3 &&  $\Delta{t}_1$ \& $\Delta{t}_3$ &  $\Delta{t}_2$ \& $\Delta{t}_3$ \\ \hline
Drag coefficient ($C_D$) & 1.0453  &  0.88467 &&    0.37801 &  0.25764 \\
Lift coefficient ($C_L$) & 1.5462 &  1.1158 & &   0.63192 &  0.4343 \\ \bottomrule \bottomrule
\end{tabular}%
}
\label{Comparison of Grid and Time-step independence study results}
\end{table}

\section{Results and Discussion}
To address the research questions outlined in Section~\ref{sec:Intro}, which focus on the coupling between vortex dynamics and the transport of chemical cues, we perform simulations of \edt{fluid flows} and odor dynamics around $\mbox{3D}$ undulating bodies. The governing flow, kinematic, and odor-related parameters used in these simulations are summarized in Table~\ref{performance_parameters}. The simulations are conducted within a transitional flow regime \cite{khalid2020flow}, with two distinct undulating waveform patterns\edt{, including} anguilliform and carangiform, over \edt{the bodies of the two swimmers} \cite{kamran2024does}. These waveforms are prescribed \edt{for} $f^\ast = 0.25$ and $0.40$, consistent with the typical swimming patterns of marine animals \cite{khalid2020flow}. For this study, we limit the governing parameters to those most relevant for $\mbox{3D}$ flows, as a detailed parametric analysis \edt{was} already performed and \edt{presented} in our recent $\mbox{2D}$ study \cite{kamran2024does}. It is \edt{important to note} that the undulation wavelength is set to $\lambda/L=0.80$ for the anguilliform kinematic mode \cite{khalid2021anguilliform} and $\lambda/L=1.05$ for the carangiform mode \cite{khalid2021larger}.

\begin{table}[htbp]
\centering
\caption{\small {Specifications of the governing parameters}}
\setlength{\tabcolsep}{50pt} 
\begin{tabular}{ll}
\hline
\hline
Parameters           & Specifications              \\ \hline
Geometry              & Eel and Jackfish                   \\
Undulatory kinematics & Anguilliform and Carangiform \\
Re                    & $3000$           \\
Sc                    & $340$ (water)    \\
$f^\ast$                 & $0.25$ and $0.4$                    \\
$\lambda$             & $0.80$ and $1.05$                \\ \hline
\hline
\end{tabular}
\label{performance_parameters}
\end{table}

We begin our discussion by presenting the vortex topology and odor dynamics around the \edt{j}ackfish with carangiform motion and the \edt{e}el with anguilliform motion at $\mbox{Re} = 3000$ and $f^\ast = 0.40$. For this purpose, we employ the Q-criterion to visualize \edt{3D} vortices \cite{hunt1988eddies}. The Q-criterion is defined as $Q = 0.5(\|{\Omega}\|^2 - \|\mathbf{S}\|^2)$, where ${\Omega}$ represents the antisymmetric part and $\mathbf{S}$ represents the symmetric part of the velocity gradient tensor. Regions with $Q > 0$, where the rotation rate dominates the strain rate, indicate distinct coherent vortices. 


\begin{figure}[htbp]
{\includegraphics[width=1.0\textwidth]{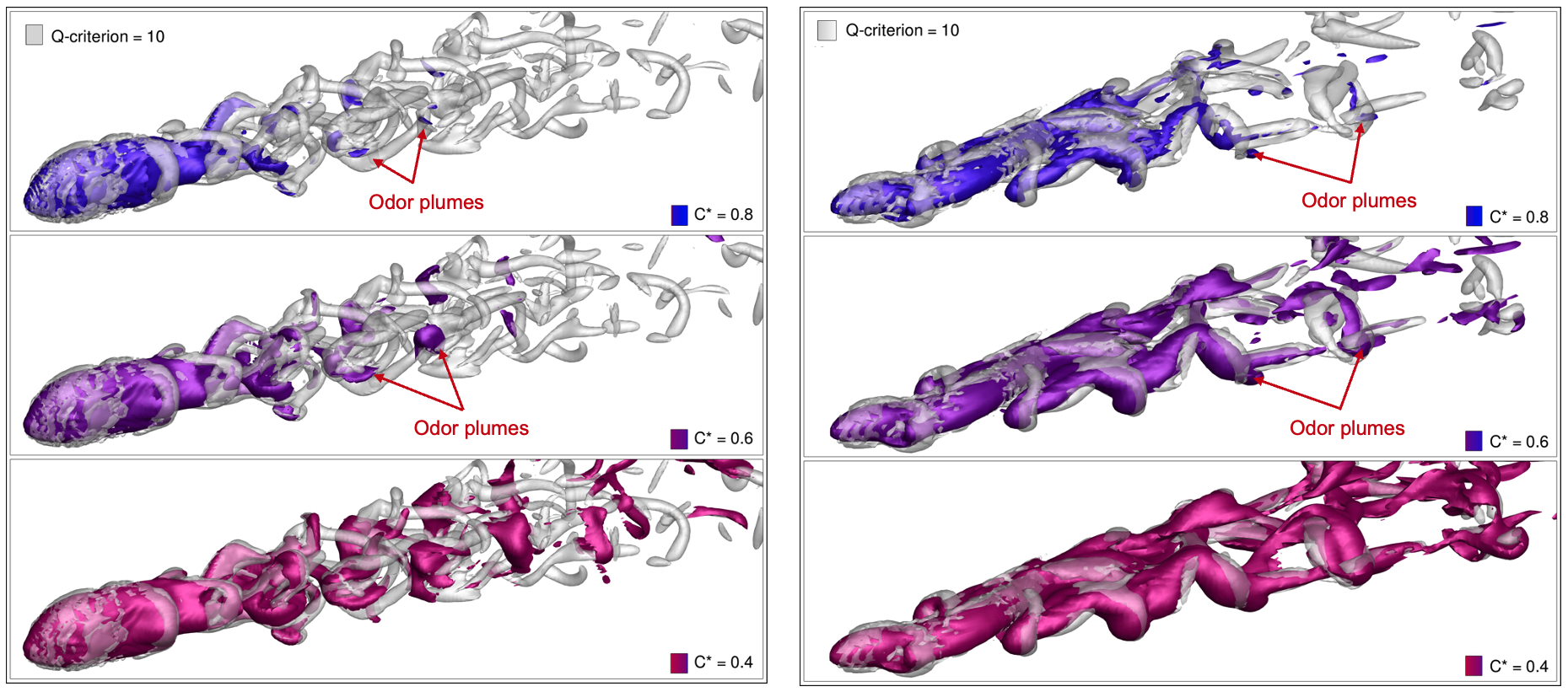}}
\caption{Vortex topology visualized using the Q-criterion with a value of $10$, at three different Odor concentration levels. The simulations are performed at \(Re = 3000\) and \(f^\ast = 0.4\). The first and second columns represent the results for \edt{j}ackfish (carangiform) and  \edt{e}el (anguilliform), respectively.}
\label{fig:odor}
\end{figure}

Figure~\ref{fig:odor} shows the Q-criterion, visualized in light grey with a value of 10, alongside the non-dimensional odor concentration ($C^*$) at three levels: $C^* = 0.8$ (blue), $C^* = 0.6$ (purple), and $C^* = 0.4$ (pink). Here, the swimmer's body serves as the source, where $C^* = 1$. Two key observations emerge from \edt{these} plots. First, the odor plumes not only \edt{strictly} follow the vortex topology but also reside within it, indicating a strong coupling between the flow structures and odor transport. Second, the \edt{e}el with anguilliform kinematics produces stronger and farther-reaching odor plumes compared to the \edt{j}ackfish with carangiform kinematics. These findings align with the results of Kamran et al. \cite{kamran2024does}, with one notable exception. \edt{T}he overall trajectory of odor spots and coherent flow structures \edt{were} not always precisely aligned in our previously reported $\mbox{2D}$ simulations, \edt{and a phase shift between the two fields was evident. This phase shift is absent in our current work as it} incorporates $\mbox{3D}$ effects, which capture more complex interactions between odor dynamics and vortical structures. 

\begin{figure}[htbp]
    \centering
    \begin{minipage}{0.5\textwidth}
        \centering
        \includegraphics[width=\linewidth]{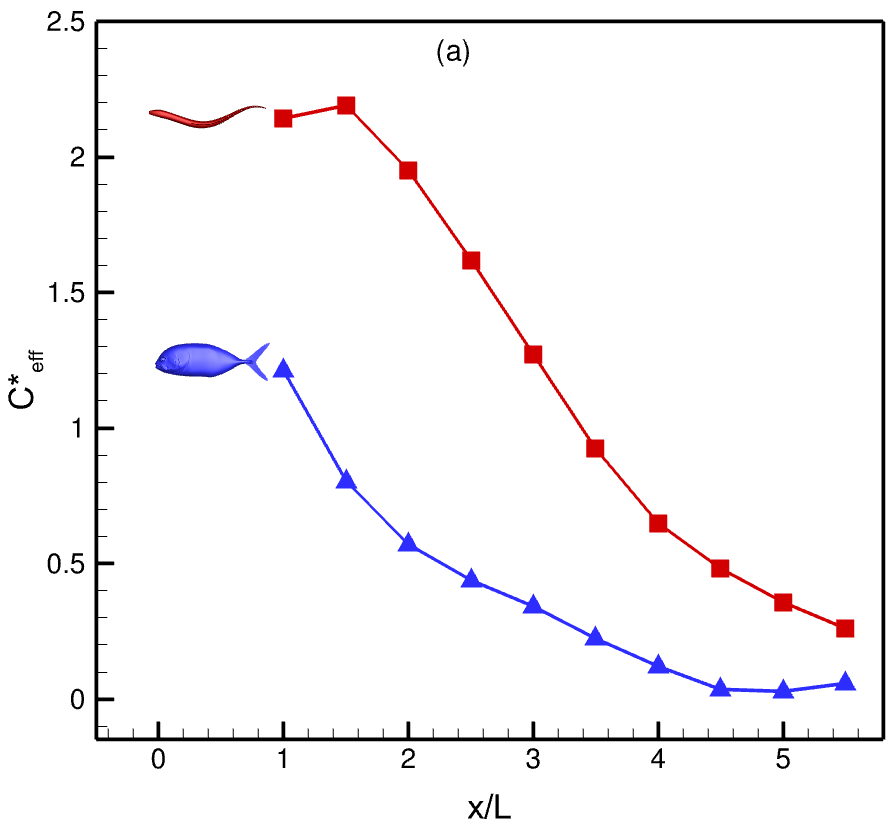}
    \end{minipage}%
    \begin{minipage}{0.5\textwidth}
        \centering
        \includegraphics[width=\linewidth]{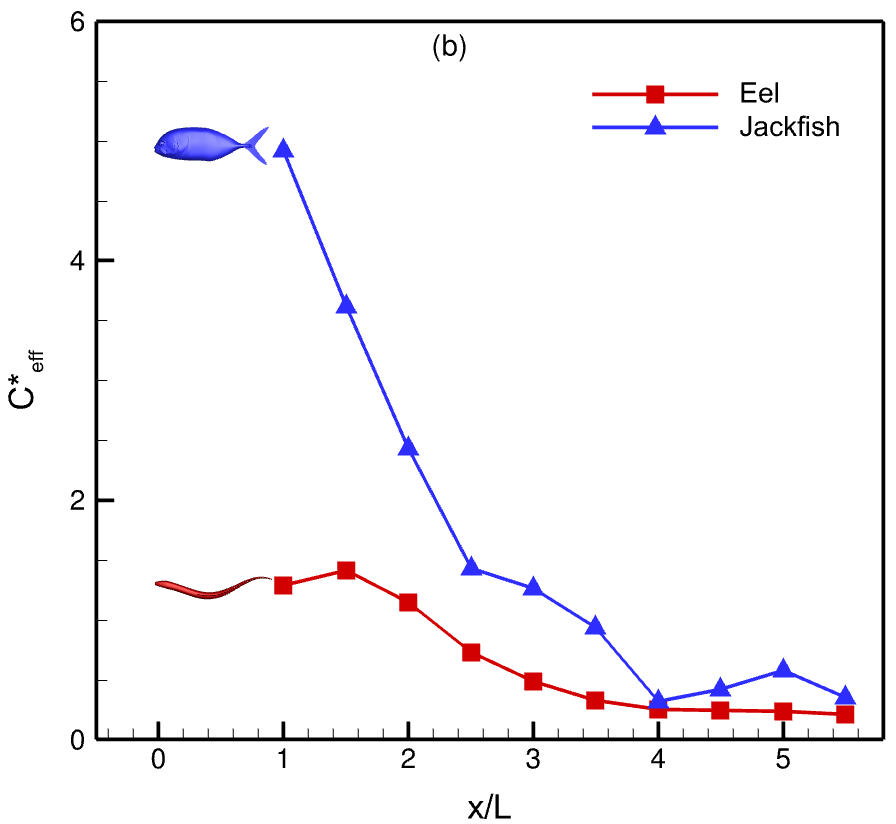}
    \end{minipage}
    \caption{Odor effectiveness in the wake of \edt{a j}ackfish and \edt{an e}el: (a) actual and (b) \edt{switched} kinematics}
    \label{fig: SA odor}
\end{figure}

To further analyze this behavior, we quantify the results presented in Fig.~\ref{fig: SA odor}. We divide the wake region, starting from the trailing edge \edt{of the swimmer}, into $10$ subzones, each with a length of $0.5L$, thereby covering the total wake region up to $5.5L$ downstream of the swimmer. For this analysis, we extract the isosurface of odor concentration at $C^* = 0.5$, as concentrations below this threshold \edt{do not} fall within the odor detection limit and have a significantly reduced effect. Data is collected at eight-time steps per oscillation cycle \edt{and time-}averaged, \edt{with the areas of the extracted isosurfaces of odor} normalized by the swimmer’s body area. This normalization allows for a direct comparison of \edt{how effectively the odor is dispersed in their wakes by the two swimmers with different undulating kinematics}.

The normalized odor effectiveness is mathematically defined as:

\begin{align}
C_{\text{eff}}^\ast = \frac{A_{\text{iso}}}{A_{\text{fish}}},
\label{eq:area_effectiveness}
\end{align}

\noindent where $A_{\text{iso}}$ is the isosurface area \edt{of odor} within a given subzone, and $A_{\text{fish}}$ is the swimmer's body area.

From Fig.~\ref{fig: SA odor}a and the observations from Fig.~\ref{fig:odor}, we note that $C_{\text{eff}}^\ast$ for the \edt{e}el with anguilliform kinematics is significantly higher than that of the \edt{j}ackfish. Although the odor source area of the jackfish is larger than that of the eel, implying a \edt{potentially} higher odor concentration in its wake, our results contradict this intuition. This discrepancy raises the question: why does the eel produce more effective odor plumes, which travel farther downstream, despite having a smaller source area? To address this, we run additional simulations in which the kinematics are swapped between the two swimmers, assigning anguilliform kinematics to the jackfish and carangiform kinematics to the eel. Following the same procedure to compute $C_{\text{eff}}^\ast$, the results are plotted in Fig.~\ref{fig: SA odor}b. These findings clarify that it is not the swimmer's morphology, but rather its kinematics, that play a dominant role in spreading odor in the wake \edt{that is an important insight for underwater sensing techniques}.

\begin{figure}[htbp]
    \centering
    \begin{minipage}{0.5\textwidth}
        \centering
        \includegraphics[width=\linewidth]{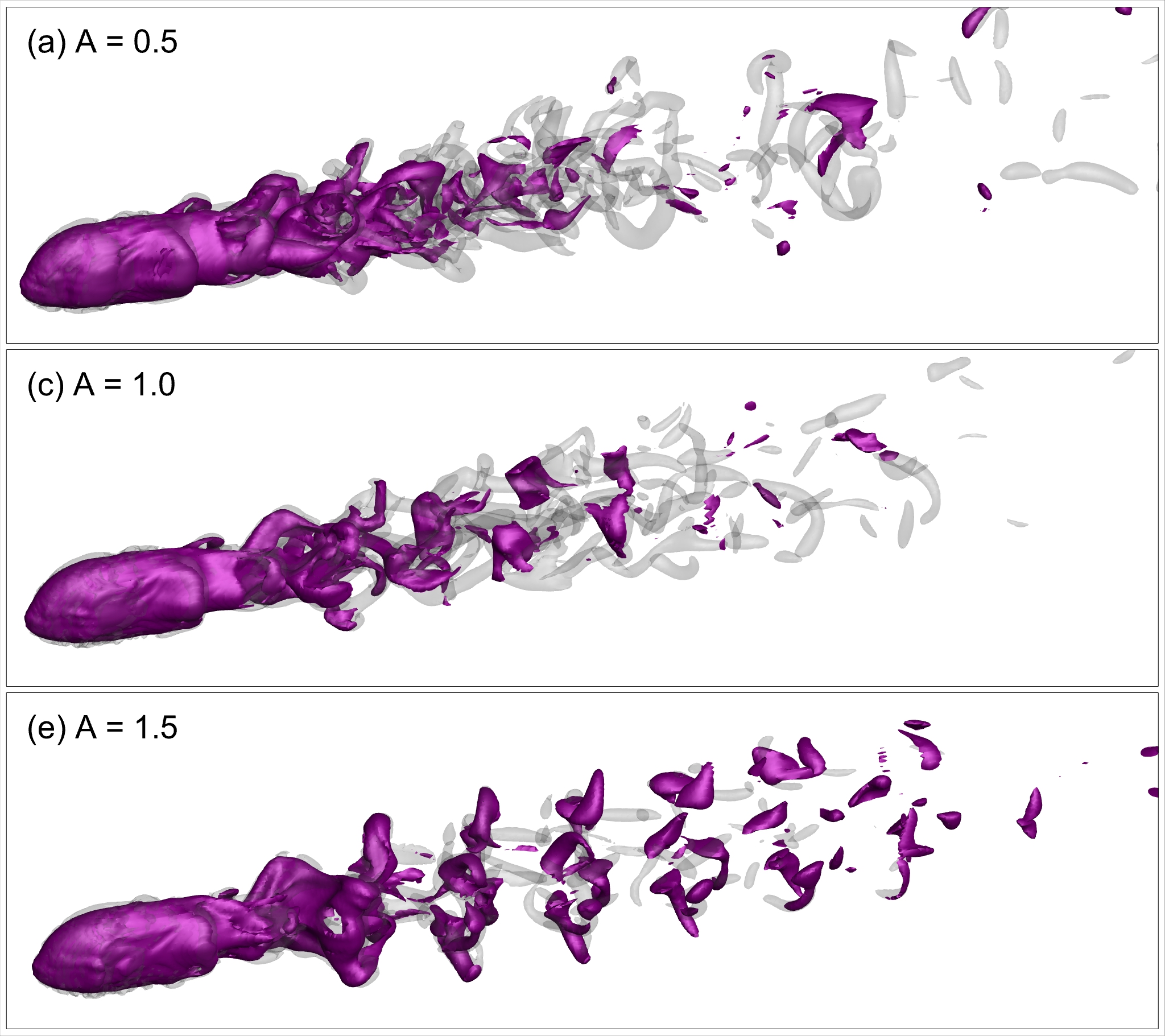}
    \end{minipage}%
    \begin{minipage}{0.5\textwidth}
        \centering
        \includegraphics[width=\linewidth]{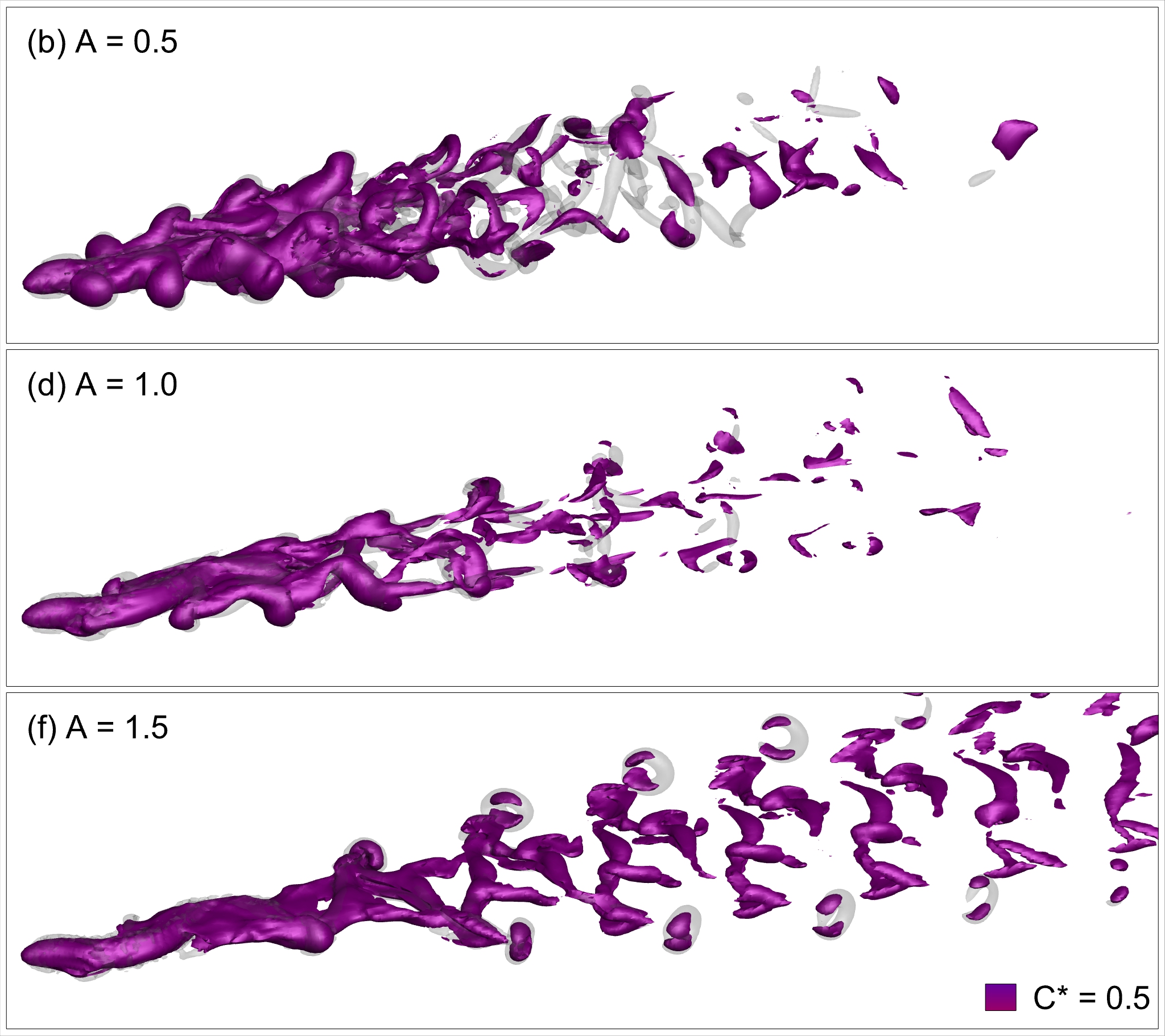}
    \end{minipage}
    \caption{Odor plumes for a jackfish (first column) and an eel (second column) with a different magnification of undulation}
    \label{fig: odor_magnify}
\end{figure}

To further strengthen our argument, we address another key research question: how do kinematics influence \edt{decay of odor in the wake}? To investigate this, we perform additional simulations by varying the magnification of undulation by $\pm 50\%$. Specifically, we adjust the amplitude profile for each swimmer's kinematics by scaling the amplitude equations for carangiform motion (Eq.~\ref{eq:carangiform}) and anguilliform motion (Eq.~\ref{eq:amp}) to achieve the desired magnification. The modified amplitude profiles are obtained by multiplying the original equations by scaling factors of $A = 0.5$ ($50\%$ decrease in magnification) and $A = 1.5$ ($50\%$ increase in magnification), while $A = 1.0$ represents the original undulation. Figure~\ref{fig: odor_magnify} \edt{exhibits} the resulting odor concentration isosurfaces at $C^* = 0.5$ for the undulatory swimmers, jackfish, and eel, at these three levels of amplitude magnification. 

\begin{figure}[htbp]
    \centering
    \begin{minipage}{0.5\textwidth}
        \centering
        \includegraphics[width=\linewidth]{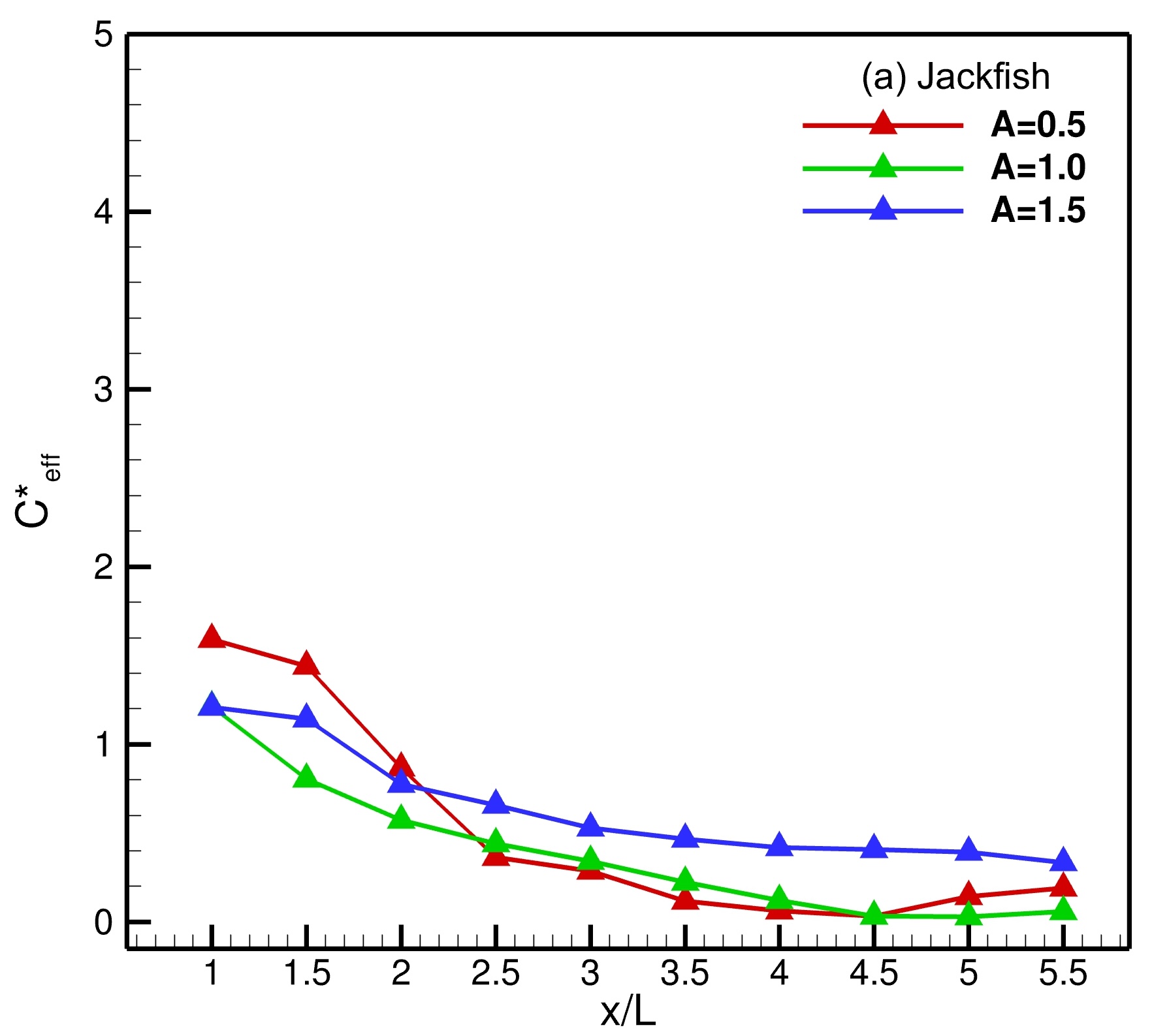}
    \end{minipage}%
    \begin{minipage}{0.5\textwidth}
        \centering
        \includegraphics[width=\linewidth]{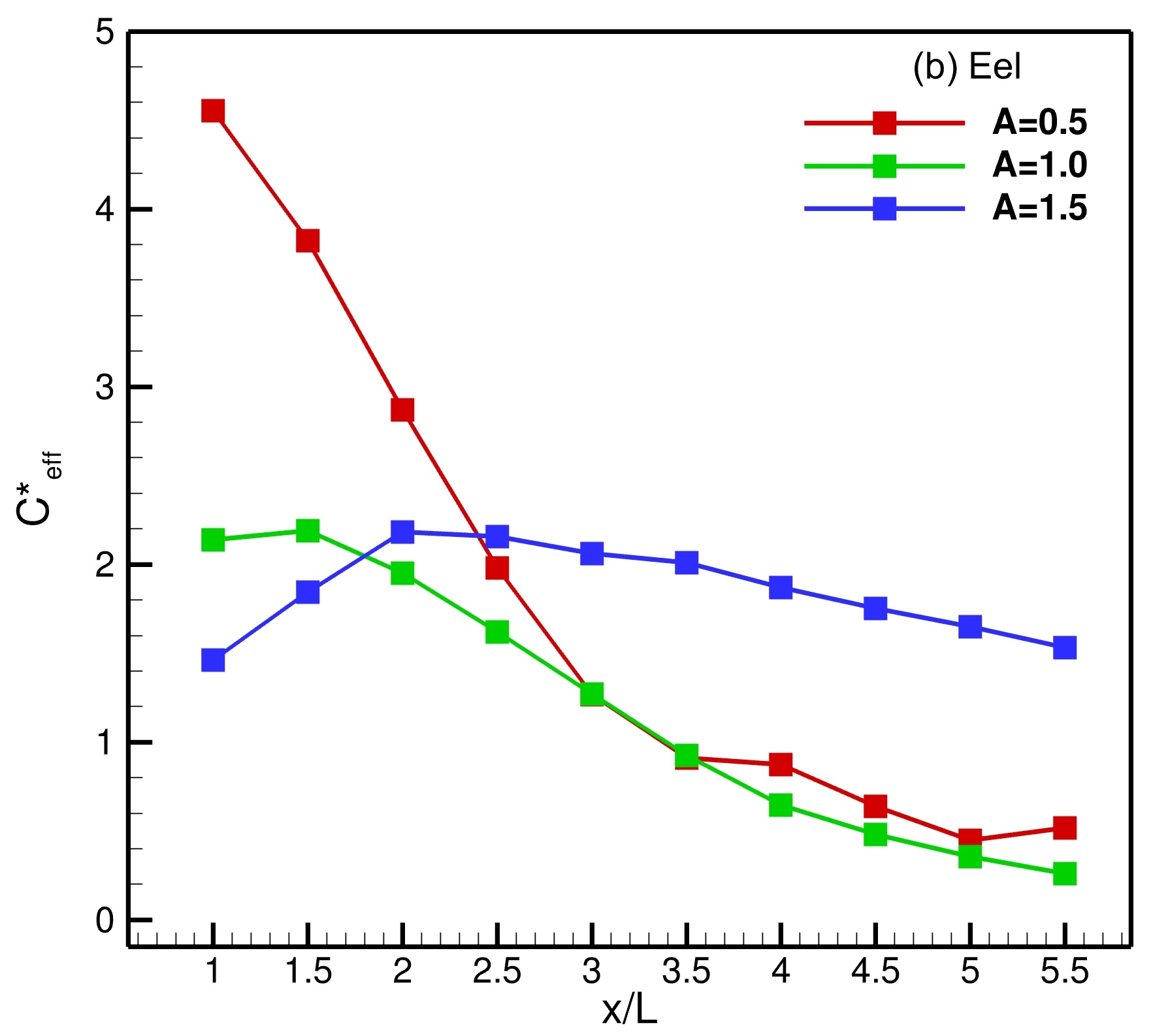}
    \end{minipage}
    \caption{Odor effectiveness with a different magnification of undulation: (a) jackfish and (b) eel}
    \label{fig: odor_magnify_eff}
\end{figure}

We observe that as the amplitude magnification increases from $A = 0.5$ to $A = 1.5$, the odor concentration becomes more intensified, indicating that vortex activity diminishes odor dispersion \edt{in the wake}. This behavior is also evident in \edt{Fig.}~\ref{fig:odor}. To quantify this effect, we \edt{compute} $C_{\text{eff}}^\ast$ for different magnifications of undulation, as shown in \edt{Fig.}~\ref{fig: odor_magnify_eff}. The results demonstrate that the odor effectiveness for the eel (\edt{Fig.}~\ref{fig: odor_magnify_eff}b) is nearly twice that of the jackfish (\edt{Fig.}~\ref{fig: odor_magnify_eff}a). Additionally, as seen in \edt{Figs.}~\ref{fig: odor_magnify} and \ref{fig: odor_magnify_eff}, the odor concentration tends to settle within a region up to $2.5L$ downstream. The differences in undulation patterns between the two swimmers further highlight the observed trends. For the jackfish, where undulation is primarily concentrated in the posterior region (mainly the tail), the odor effectiveness decreases exponentially in the wake. In contrast, for the eel, which exhibits undulation along its entire body, the maximum odor effectiveness at $A = 0.5$ and $A = 1.0$ follows an exponential decay, whereas at $A = 1.5$, the decay \edt{starts occurring at a} farther downstream \edt{location}, approximately at \edt{a distance of} $2L$ \edt{from the swimmer}. Finally, the maximum odor effectiveness for both swimmers is observed at $A = 1.5$. This is because the increased magnification of undulation amplitude leads to greater work done on the surrounding fluid, thereby increasing momentum transfer and enhancing odor transport in the wake.

\begin{figure}[htbp]
    \centering
    \begin{minipage}{0.5\textwidth}
        \centering
        \includegraphics[width=\linewidth]{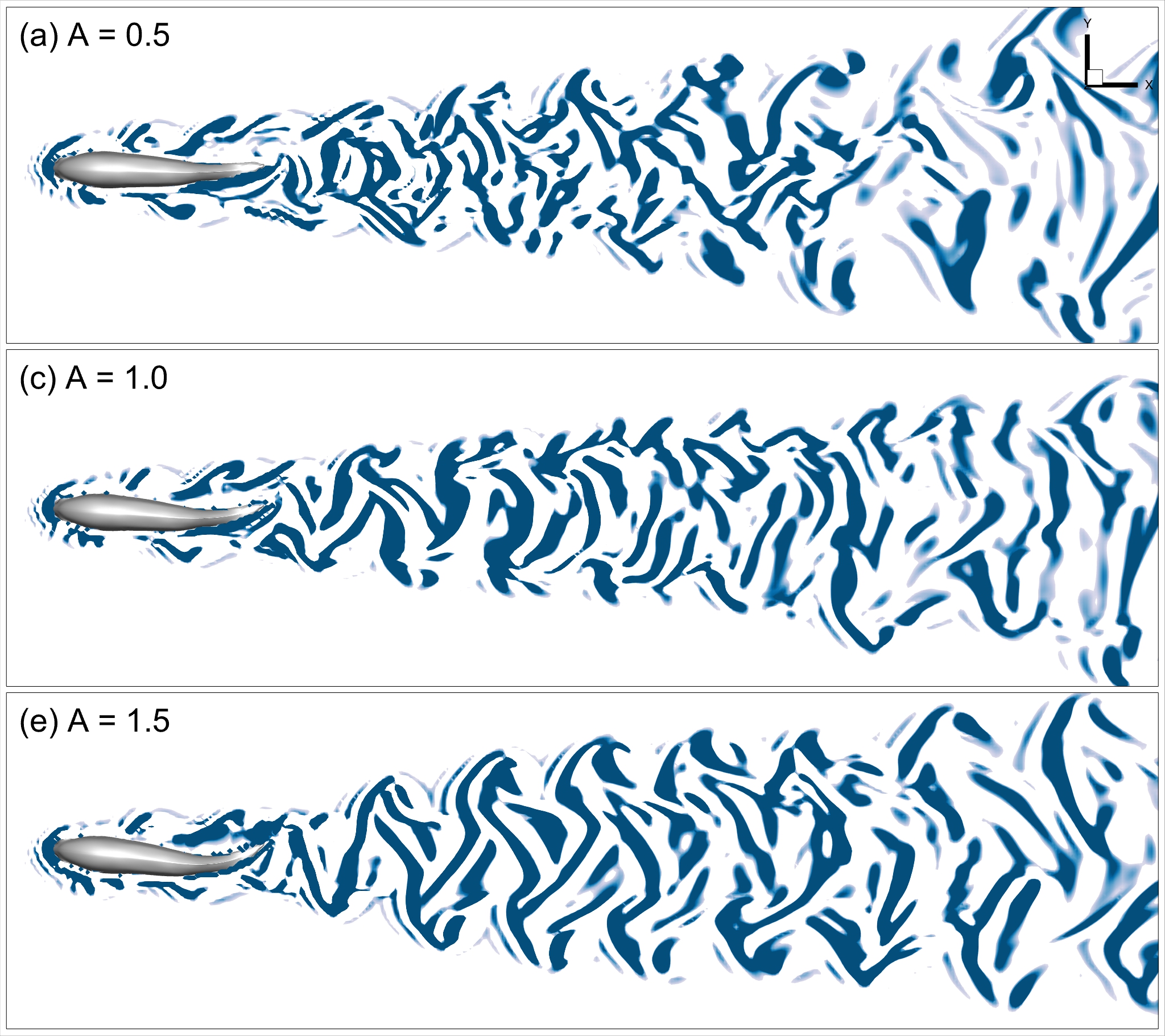}
    \end{minipage}%
    \begin{minipage}{0.5\textwidth}
        \centering
        \includegraphics[width=\linewidth]{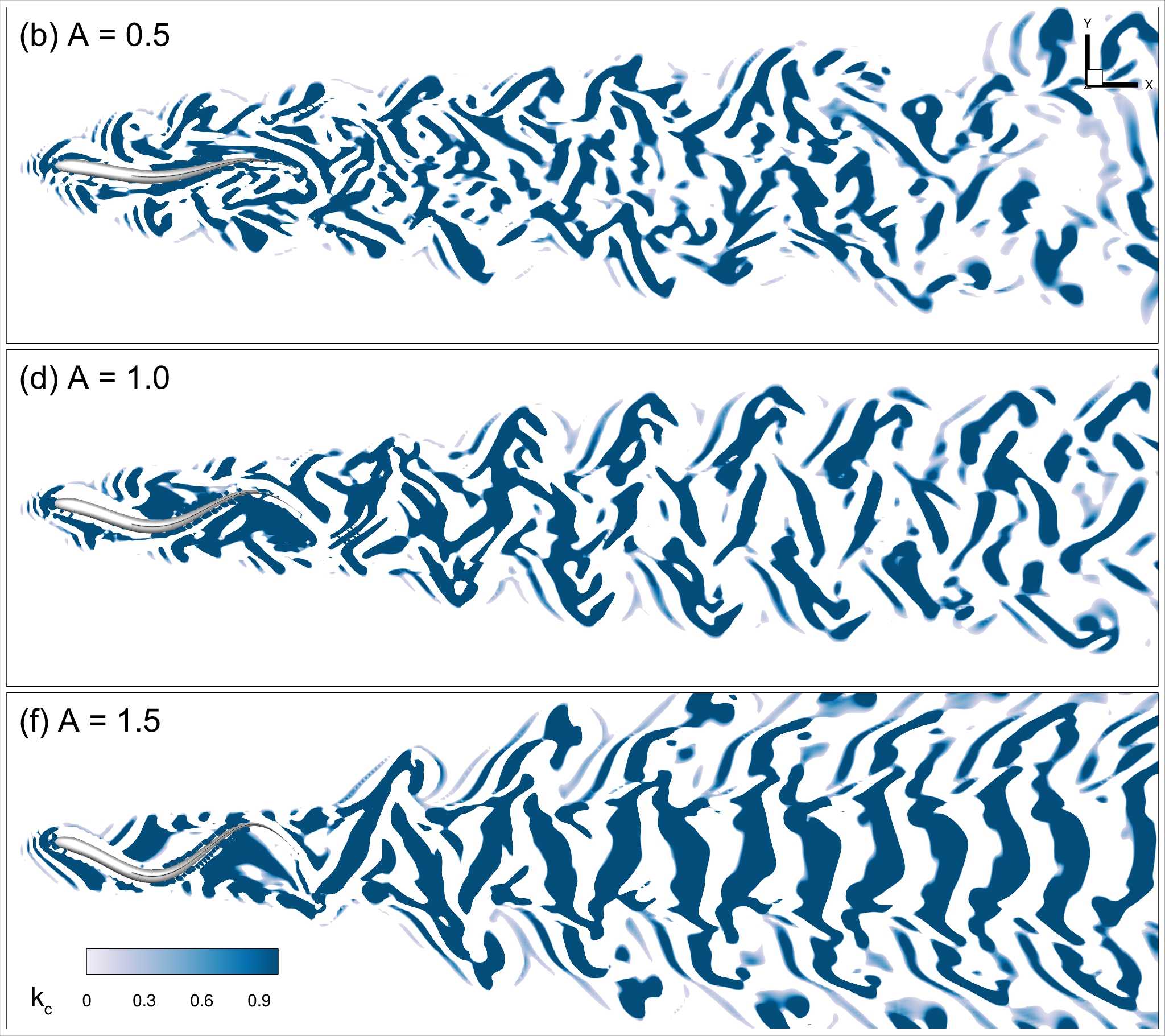}
    \end{minipage}
    \caption{Influence of convection for a jackfish (first column) and an eel (second column) with a different magnification of undulation}
    \label{fig: convection}
\end{figure}

\begin{figure}[htbp]
    \centering
    \begin{minipage}{0.5\textwidth}
        \centering
        \includegraphics[width=\linewidth]{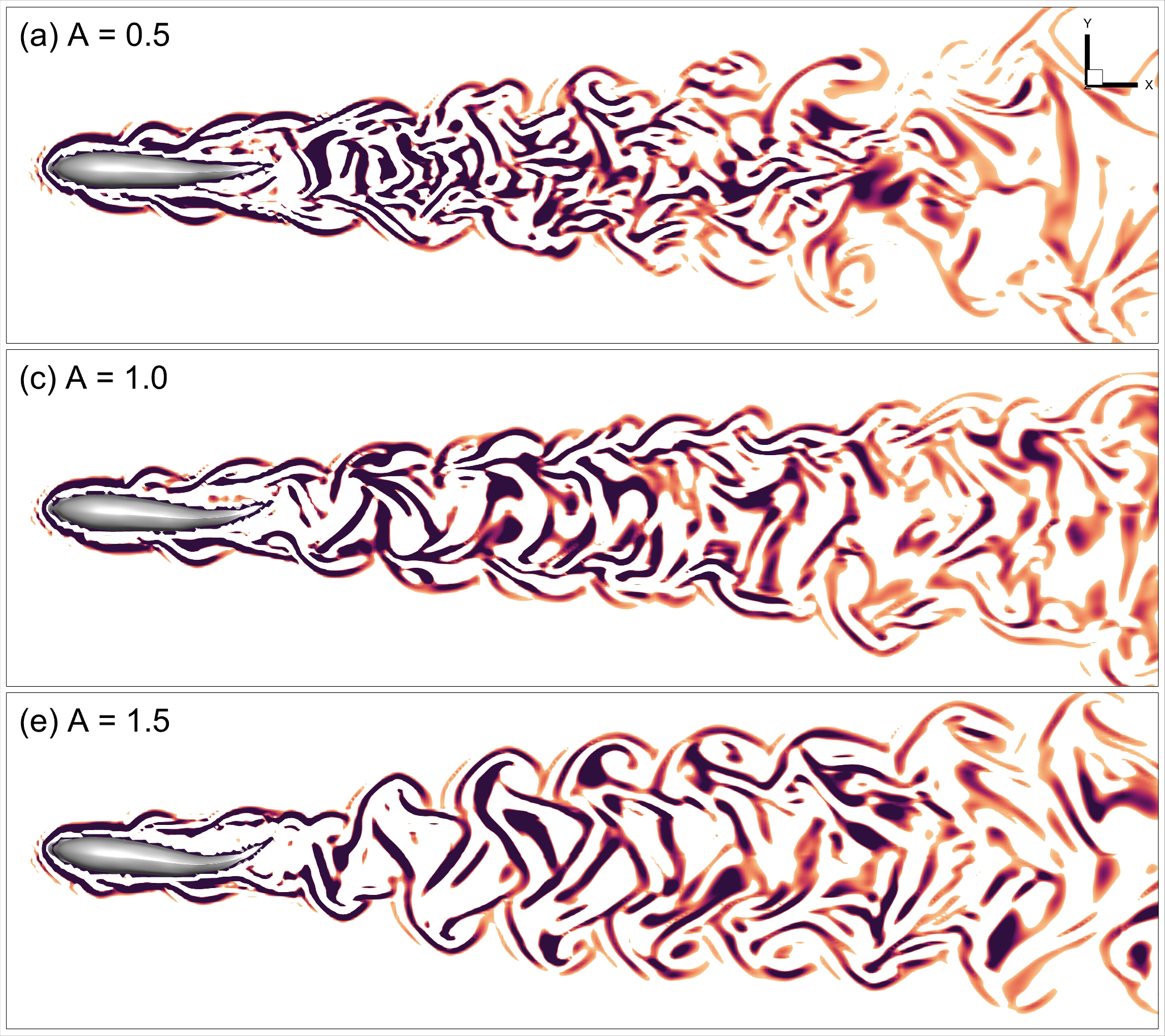}
    \end{minipage}%
    \begin{minipage}{0.5\textwidth}
        \centering
        \includegraphics[width=\linewidth]{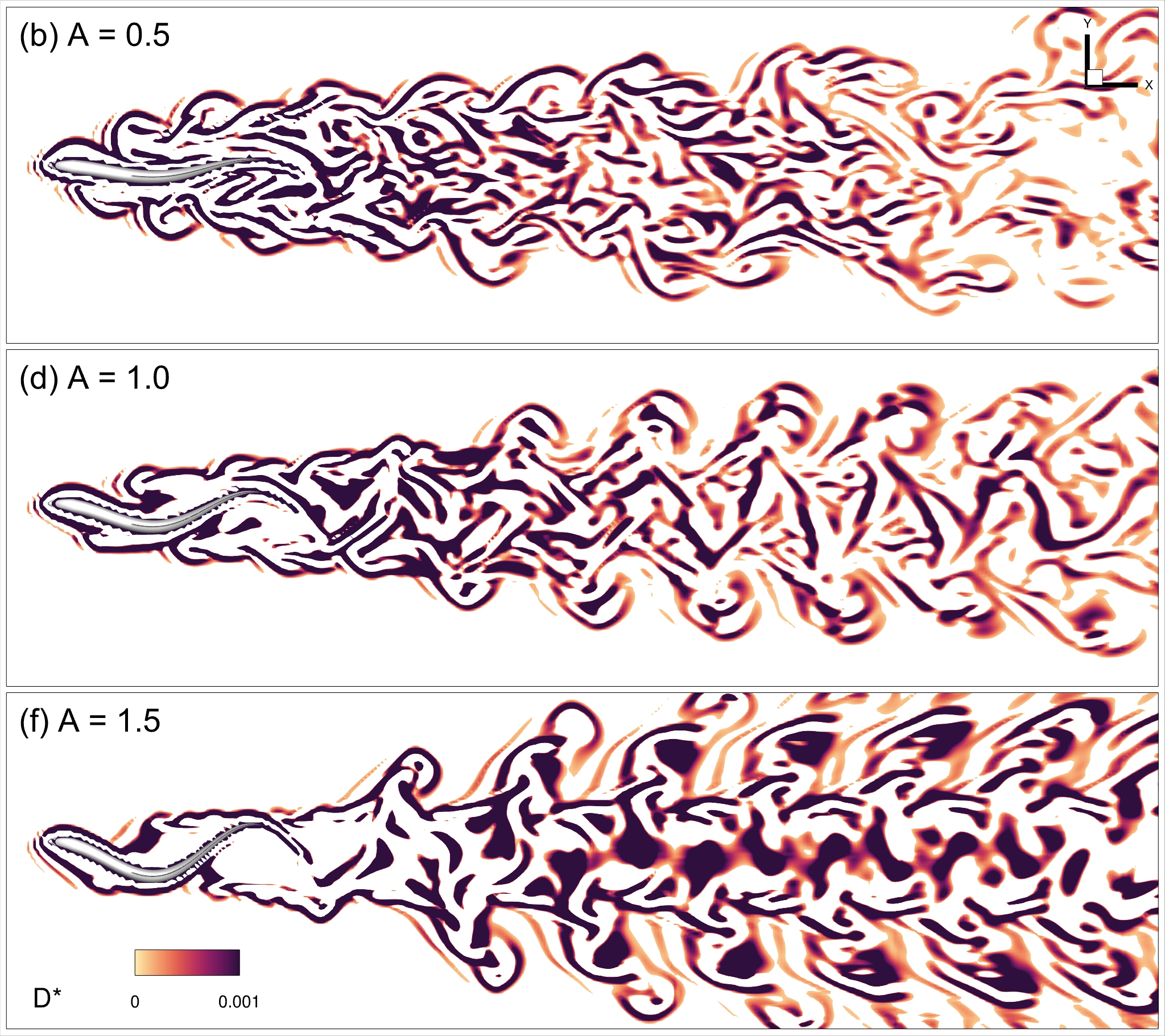}
    \end{minipage}
    \caption{Influence of diffusion for a jackfish (first column) and an eel (second column) with a different magnification of undulation}
    \label{fig: diffusion}
\end{figure}

An important aspect of this study is to determine the influence of convection and diffusion \edt{components} on the transport of chemical cues in the wake \cite{kamran2024does}. To evaluate these effects individually, we discretize the advection-diffusion equation (Eq.~\ref{eqn:odor_transport}) and analyze the contributions of the convection and diffusion terms separately. The convection term (the second term on the left-hand side) is discretized using a QUICK scheme to ensure stability in the presence of advection-dominated flows, while the diffusion term (on the right-hand side) is discretized using \edt{the} Crank\edt{-}Nicolson scheme to capture the finer gradients of diffusive transport accurately. To gain a more detailed understanding of these mechanisms, we plot top views and $z$-axis slices \edt{(mid planes)} of the flow fields for both convection and diffusion, as shown in \edt{Figs.}~\ref{fig: convection} and \ref{fig: diffusion}. These visualizations enable us to examine how chemical cues spread through the wake region under the influence of these two processes. 

At a 50\% increased magnification of undulation ($A = 1.5$), convection ($k_c$) significantly dominates the wake dynamics for both the jackfish and eel, leading to an intensified odor concentration in the downstream region. The amplified motion results in higher momentum transfer to the surrounding fluid, which in turn accelerates the convective transport of odor.  This observation aligns with \edt{the results in Figs.}~\ref{fig: odor_magnify} and \ref{fig: odor_magnify_eff}, where the stronger undulatory motion enhances odor transport in the wake. \edt{On the contrary}, at a lower undulation magnitude ($A = 0.5$), the reduced body motion leads to lower momentum transfer to the surrounding fluid. As a result, convection plays a less dominant role, and diffusion ($D^*$) becomes more significant, particularly in the cross\edt{-}flow direction. \edt{It} means that, instead of being advected \edt{effectively in the downstream direction}, the odor disperses more laterally due to molecular diffusion. This diffusive transport causes the chemical cues to spread transversely rather than forming a concentrated wake structure, leading to a lower net odor transport downstream.

A noteworthy observation is that while diffusion plays a role in transverse dispersion, its overall effect is minimal compared to convection. \edt{It} is primarily due to the high Schmidt number for water, which indicates that the relative importance of molecular diffusion is much smaller than that of convection in aquatic environments. \edt{Consequently}, the primary transport mechanism for odor in the wake of \edt{the} undulating swimmers is dominated by convective processes, especially at higher undulation amplitudes, while diffusion provides a secondary effect, primarily contributing to the lateral spread of odor.

\section{Conclusions}
\mk{This study explores the intricate relationship between vortex dynamics and odor transport in the wake of biological swimmers, emphasizing the impact of kinematics and morphology on chemical dispersion. Using high-fidelity three-dimensional simulations, we analyze the hydrodynamic and chemical wake characteristics of carangiform (jackfish) and anguilliform (eel) swimmers. Our findings reveal that kinematics, rather than the physiological shape, predominantly dictates odor dispersal, with anguilliform swimmers generating broader and more persistent odor trails. To further investigate this effect, we swap the kinematics between the two swimmers, assigning anguilliform motion to a jackfish and carangiform motion to an eel. The results confirm that it is not the swimmer’s morphology, but the kinematics that plays a dominant role in spreading odor in the wake, which is an important insight to understand underwater sensing techniques. Additionally, we find that increasing the undulation amplitude enhances odor transport, as greater magnification of the undulation leads to increased work done on the surrounding fluid, thereby increasing momentum transfer and enhancing spread of odor in the wake. The strong coupling between vortex structures and chemical cues highlights the dominance of convection over diffusion in aquatic environments. These insights not only enhance our understanding of underwater sensory mechanisms in marine species but also provide valuable guidance for designing bio-inspired robotic systems with enhanced chemical detection and navigation capabilities. By bridging the gap between fluid dynamics and chemical transport, this work contributes to advancements in marine biology, environmental conservation, and autonomous underwater robotics.}



\section*{Acknowledgment}
MSU Khalid acknowledges funding support from the Natural Sciences and Engineering Research Council of Canada (NSERC) through the Discovery and Alliance International grant programs for this work. C. Li acknowledges funding support from the National Science Foundation (CBET-2453175) monitored by Dr. R. D. Joslin and the Air Force Office of Scientific Research (FA9550-24-1-0122) monitored by Dr. Patrick Bradshaw.

\vskip6pt

\bibliography{Manuscript}

\end{document}